\documentstyle[12pt]{article}
\begin{document}
\title{A Reconciliation of Electromagnetism and Gravitation}
\author{B.G.Sidharth \\
Centr for Applicable Mathematics and Computer Sciences\\
B.M.Birla Science Centre,Hyderabad, 500063,India}
\date{}
\maketitle
\begin{abstract}
It is argued that once we consider the underpinning of a Non Commutative
geometry, itself symptomatic of extended particles, for example in Quantum
Superstring theory, then a reconciliation between gravitation and electromagnetism
is possible.
\end{abstract}
\section{Introduction}
Despite nearly a century of work, it has not been possible to achieve a
unification of gravitation and electromagnetism. It must be borne in mind
that the tools used, be it Quantum Theory or General Relativity are deeply
entrenched in differentiable space time manifolds (and point particles) - the former with Minkowski
space time and the latter with curved space time. The challenge has been, as
Wheeler noted\cite{r1}, the introduction of Quantum Mechanical spin half into
General Relativity on the one hand and the introduction of curvature into
Quantum Mechanics on the other.\\
More recent models including Quantum Superstrings on the contrary deal with
extended and not point particles and lead to a non-differentiable spacetime
and a non commutative geometry (NCG)\cite{r2,r3,r4,r5}.\\
Indeed way back in the 1930s, Einstein himself observed\cite{r6}
"...it has ben pointed out that the
introduction of a space-time continuum may be considered as contrary to nature in
view of the molecular structure of everything which happens on a small
scale. It is maintained that perhaps the success of the Heisenberg method points
to a purely algebraic method of description of nature that is to the elimination
of continuous functions from physics. Then however, we must also give up, by
principle the space-time continuum. It is not unimaginable that human ingenuity
will some day find methods which will make it possible to proceed along such a
path."\\
Even at the beginning of the twentieth century several Physicists including Poincare
and Abraham amongst others were working unsuccessfully with the problem of
the extended electron\cite{r7,r8}. The problem was that an extended electron
appeared to contradict Special Relativity, while on the other hand, the limit
of a point particle lead to inexplicable infinities. These infinities dogged
physics for many decades. Infact the Heisenberg Uncertainity Principle straightaway
leads to infinities in the limit of spacetime points. It was only through
the artifice of renormalization that 't Hooft could finally circumvent this
vexing problem, in the 1970s\cite{r9}.\\
Nevertheless it has been realized that the concept of spacetime points is only
approximate. We are beginning to realize that it may be
more meaningful to speak in terms of spacetime foam, strings, branes, non
commutative geometry, fuzzy spacetime and so on\cite{r10}.\\
Indeed non commutativity arises if there is a minimum space time
length as shown a long time ago by Snyder\cite{r11}. What we will argue below
is that once the underlying non commutative nature of the geometry is
recognized then it is possible to reconcile electromagnetism and gravitation.
\section{NCG}
It is well known that once we consider non zero minimum space time intervals
or equivalently extended particles as in Quantum Superstrings, then we have
the following non commutative geometry (Cf.refs.\cite{r2}-\cite{r5},\cite{r11}):
\begin{equation}
[x,y] = 0(l^2), [p_x,p_y] \approx \frac{\hbar^2 0(1)}{l^2}\label{e1}
\end{equation}
(and similar equations)
where $l,\tau$ are the extensions of the space time coordinates.\\
In conventional theory the space time coordinates as also the momenta commute
amongst themselves unlike in equation (\ref{e1}). It must be observed that the
non commutative relations are self evident, in the sense that $xy$ or $yx$ is
each of the order of $l^2$, and so is their difference because of the non
commutativity.\\
The non commutative or in Witten's words, Fermionic feature is symptomatic of the breakdown
of the concept of the spacetime points and point particles at small scales
or high energies. As has been noted by Snyder, Witten, and several other
scholars, the divergences encountered in Quantum Field Theory are symptomatic
of precisely such an extrapolation to spacetime points and which necessitates
devices like renormalization. As Witten points out\cite{r12}, "in developing
relativity, Einstein assumed that the space time coordinates were Bosonic;
Fermions had not yet been discovered!... The structure of space time is
enriched by Fermionic as well as Bosonic coordinates."\\
Interestingly it has been shown that the commutative relations (\ref{e1})
lead directly to the Dirac equation, on the one hand\cite{r13}. On the other hand,
it is interesting that a differential calculus over a non commutative algebra
uniquely determines a gravitational field in the commutative limit and that there
is a unique metric which remains as a classical "shadow" as shown by Madore\cite{r14}.\\
Let us now introduce this effect into the usual distance formula in flat space
\begin{equation}
ds^2 = g_{\mu \nu} dx^\mu dx^\nu\label{e2}
\end{equation}
Rewriting the product of the two coordinate differentials in (\ref{e2}) in
terms of the symmetric and non symmetric combinations, we get
\begin{equation}
g_{\mu \nu} = \eta_{\mu \nu} + kh_{\mu \nu}\label{e3}
\end{equation}
where the first term on the right side of (\ref{e3}) denotes the usual flat space
time and the second term denotes the effect of the non commutativity, $k$ being
a suitable constant.\\
It must be noted that if $l,\tau \to 0$ then equations (\ref{e1}) and also (\ref{e3})
reduce to the usual formulation.\\
The effect of the non commutative geometry is therefore to introduce a departure
from flat space time, as can be seen from (\ref{e3}).\\
Infact remembering that the second term of the right side of (\ref{e3}) is small,
this can straightaway be seen to lead to a linearized theory of General
Relativity\cite{r15}. Exactly as in this reference we could now deduce the
General Relativistic relation
$$\partial_\lambda \partial^\lambda h^{\mu \nu} - (\partial_\lambda \partial^\nu
h^{\mu \lambda} + \partial_\lambda \partial^\mu h^{\nu \lambda})$$
\begin{equation}
-\eta^{\mu \nu} \partial_\lambda \partial^\lambda h + \eta^{\mu \nu} \partial_\lambda
\partial_\sigma h^{\lambda \sigma} = - kT^{\mu \nu}\label{e4}
\end{equation}
Let us now consider the non commutative relation (\ref{e1}) for the momentum
components. Then, it can be shown using (\ref{e1}) and (\ref{e3}) that\cite{r16},
\begin{equation}
\frac{\partial}{\partial x^\lambda} \frac{\partial}{\partial x^\mu} -
\frac{\partial}{\partial x^\mu} \frac{\partial}{\partial x^\lambda} \quad \mbox{goes}
\quad \mbox{over}\quad \mbox{to} \frac{\partial}{\partial x^\lambda}
\Gamma^\nu_{\mu \nu} - \frac{\partial}
{\partial x^\mu} \Gamma^\nu_{\lambda \nu}\label{e5}
\end{equation}
Normally in conventional theory the right side of (\ref{e5}) would vanish. Let
us designate this nonvanishing part on the right by
\begin{equation}
\frac{e}{c\hbar} F^{\mu \lambda}\label{e6}
\end{equation}
(\ref{e6}) can be written as
\begin{equation}
Bl^2 \sim \frac{\hbar c}{e}\label{e7}
\end{equation}
where $B$ is the magnetic field, if we are to identify $F^{\mu \nu}$ with the electromagnetic
tensor\cite{r16}. It will be recognized that (\ref{e7}) gives the celebrated
expression for the magnetic monopole, and indeed it has also being shown that
a non commutative space time at the extreme scale throws up the monopole\cite{r17,r18}.\\
We have shown here that the non commutativity in momentum components leads
to an effect that can be identified with electromagnetism and infact from
expression (\ref{e6}) we have
\begin{equation}
A^{\mu} = \hbar \Gamma^{\mu \nu}_\nu\label{e8}
\end{equation}
where $A_\mu$ is the electromagnetic four potential.\\
Thus non commutativity as expressed in equations (\ref{e1}) generates both
gravitation and electromagnetism.\\
We can see this in greater detail as follows. The gravitational field equations
can be written as\cite{r15}
\begin{equation}
\Box \phi^{\mu \nu} = - kT^{\mu \nu}\label{e55}
\end{equation}
where
\begin{equation}
\phi^{\mu \nu} = h^{\mu \nu} - \frac{1}{2} \eta^{\mu \nu} h\label{e54}
\end{equation}
It also follows, if we use the usual gauge and equation (\ref{e8}) that
\begin{equation}
\partial_{\mu} h^{\mu \nu} = A^{\nu}\label{ex}
\end{equation}
in this linearised theory.\\
Whence, remembering that we have (\ref{e3}), operating on both sides of equation
(\ref{e55}) with $\partial_\mu$ we get Maxwell's equations of electromagnetism.\\
Indeed this is not surprising because, as is well known equation (\ref{e8})
is mathematically identical to the formulation of Weyl\cite{r19}.
However in Weyl's formulation, the electromagnetic potential was put in by hand.
In the above case it is a consequeence of the non commutative geometry at
small scales, which again is symptomatic of the spinorial behaviour of the
electron, as has been discussed in detail elsewhere and infact (\ref{e8})
deduced from an alternative viewpoint\cite{r5,r16}.
It is also well known that if equation (\ref{e8}) holds then in the absence
of matter th general relativistic field equations (\ref{e4}) reduce to Maxwell
equations\cite{r20}. In any case, all this provides a rationale for the fact
that from (\ref{e55}) we get the equation for spin 2 gravitons (Cf.ref.\cite{r15}) while
from the Maxwell equations, we have Spin 1 (vector) photons.
\section{Discussion}
1. The characterization of the metric in equations (\ref{e2})
and (\ref{e3}) in terms of symmetric and non symmetric components is similar
to the torsional formulation of General Relativity\cite{r21}. However in this
latter case, there is no contribution to the differential interval from the
torsional (that is non-commutative) effects. The non-commutative contribution
is given by (\ref{e1}) and herein comes the extended, rather than point like
particle.\\
In any case the above attempt at unification of electromagnetism and
gravitation had made part headway, but unless the underpinning of a non
commutative geometry is recognised, the full significance does not manifest
itself.\\
2. We now make the following remarks:\\
It can be seen from the transition to (\ref{e3}) from (\ref{e2}), that the
curvature arises from the non commutativity of the coordinates. Indeed this is
the classical analogue of a Quantum Mechanical result deduced earlier that the
origin of mass is in the minimum space intervals and the non local Quantum
Mechanical amplitudes within them as has been discussed in detail in references
cited\cite{r22,r23}. In Quantum Superstring theory also, the mass arises out
of the tension of the string in this minimum interval. We see here the convergence of the
Quantum Mechanical and classical approaches once the extension of particles is
recognized.\\
We also know that the minimum space time intervals are at the Compton scale where
the momentum $p$ equals $mc$. For a Planck mass $\sim 10^{-5}gms$, this is
also the Planck scale, as in Quantum Superstring theory.\\
In Snyder's original work, the commutation relations like (\ref{e1}) hold good
outside the minimum space time intervals, and are Lorentz invariant. This is
quite pleasing because in any case, even in Quantum Field Theory, we use
Minkowski space time.\\
3. The above non commutative geometry also holds the key to the mysterious extra
dimensions of Quantum Superstrings. This has been discussed in detail in
references\cite{r5,r18}. But to see in a simple way, we note that
equation (\ref{e1}) shows that the coordinates $y$ and $z$ show up as some sort
of a momenta, though with a different multiplying constant as the analogue of
the Planck constant. That is instead of the single $x$ momentum, $p_x$, we have
two extra momenta, this being the same for the $y$ and $z$ momenta also. This
leads to the well known $9 + 1$ dimensions of Quantum superstrings, though
because for all these extra "momenta", the multiplying factor, the analogue
of the Planck constant is different, so these extra dimensions are
supressed or curled up in the Kaluza-Klein sense.\\
4. A concept which one encounters in Quantum SuperString theory and more generally
in the presence of the Non commutative geometry (\ref{e1}) is that of Duality.
We will briefly examine this now and see its significance in relation to electrodynamic
theory.
Infact the relation (\ref{e1}) leads to\cite{r18},
\begin{equation}
\Delta x \sim \frac{\hbar}{\Delta p} + \alpha' \frac{\Delta p}{\hbar}\label{e18}
\end{equation}
where $\alpha' = l^2$, which in Quantum SuperStrings Theory $\sim 10^{-66}$.
This is an expression of the duality relation,
$$R \to \alpha'/R$$
This is symptomatic of the fact that we cannot go down to arbitrarily small
spacetime intervals, below the Planck scale in this case (Cf.ref.\cite{r24}).\\
There is an interesting meaning to the duality relation arising from (\ref{e18}).
While it appears that the ultra small is a gateway to the macro cosmos, we
could look at it in the following manner. The first term of the relation
(\ref{e18}) which is the usual Heisenberg Uncertainity relation is supplemented
by the second term which refers to the macro cosmos.\\
Let us consider the second term in (\ref{e18}). We write $\Delta p = \Delta Nmc$,
where $\Delta N$ is the Uncertainity in the number of particles, $N$, in the
universe. Also $\Delta x = R$, the radius of the universe which $\sim \sqrt{N}l$,
the famous Eddington relationship. It should be stressed that the otherwise
emperical Eddington formula, arises quite naturally in a Brownian characterisation
of the universe as has been pointed out earlier (Cf. for example ref.\cite{r25}).\\
We now get,
$$\Delta N = \sqrt{N}$$
Substituting this in the time analogue of the second term of (\ref{e18}), we
immediately get, $T$ being the age of the universe,
$$T = \sqrt{N} \tau$$
In the above analysis, including the Eddington formula, $l$ and $\tau$ are the
Compton wavelength and Compton time of a typical elementary particle, namely
the pion. The equation for the age of the universe is also correctly given
above. Infact in the closely related model of fluctuational cosmology (Cf. for
example ref.\cite{r23}) all of the Dirac large number coincidences including the
above Eddington formula as also
the mysterious Weinberg formula relating the mass of the pion to the Hubble
constant, follow as a consequence, and are not emperical. All these relations
relating large scale parameters to microphysical constants were shown to be
symptomatic of what has been called, stochastic holism (Cf. also ref.\cite{r5}),
that is a micro-macro connection with a Brownian or stochastic underpinning.
Duality, or equivalently, relation (\ref{e18}) is really an expression of
this micro-macro link.\\
We will now see a curious connection between the foregoing analysis with the apparently
disparate concept of the Feynman-Wheeler action at a distance theory, which had been quite
successful.\\
Our starting point is the so called Lorentz-Dirac equation\cite{r8}:
\begin{equation}
ma^\mu = F^\mu_{in} + F^\mu _{ext} + \Gamma^\mu\label{e101}
\end {equation}
where
$$F^\mu_{in} = \frac{e}{c} F^{\mu v}_{in} v_v$$
and $\Gamma^\mu$ is the Abraham radiation reaction four vector related to
the self force and, given by
\begin{equation}
\Gamma^\mu = \frac{2}{3} \frac{e^2}{c^3} (\dot a^\mu - \frac{1}{c^2} a^\lambda
a_\lambda v^\mu)\label{e102}
\end{equation}
Equation (\ref{e101}) is the relativistic generalisation for a point electron of
an earlier equation proposed by Lorentz, while equation (\ref{e102}) is the
relativisitic generalisation of the original radiation reaction term due to
energy loss by radiation. It must be mentioned that the mass $m$ in equation
(\ref{e101}) consists of a neutral mass and the original electromagnetic mass
of Lorentz, which latter does tend to infinity as the electron shrinks to a
point, but, this is absorbed into the neutral mass. Thus we have the forerunner
of renormalisation in quantum theory.\\
There are three unsatisfactory features of the Lorentz-Dirac
equation (\ref{e101}).\\
Firstly the third derivative of the position coordinate in (\ref{e101})
through $\Gamma^\mu$ gives
a whole family of solutions. Except one, the rest of the solutions are run away -
that is the velocity of the electron increases with time to the velocity of
light, even in the absence of any forces. This energy can be thought to come
from the infinite self energy we get when the size of the electron shrinks
to zero. If we assume adhoc an
asymptotically vanishing acceleration then we get a physically meaningful
solution, though this leads to a second difficulty,
that of violation of causality of even the physically
meaningful solutions.\\
It has been shown in detail elsewhere\cite{r7} that these acausal, non local
effects take place within the Compton time.\\
We now come to the Feynman-Wheeler action at a distance theory\cite{r26,r27}.
They showed that the apparent acausality of the theory would disappear if the
interaction of a charge with all other charges in the universe, such that the
remaining charges would absorb all local electromagnetic influences was
considered. The rationale behind this was that in an action at a distance
context, the motion of a charge would instantaneously affect other charges,
whose motion in turn would instantaneously affect the original charge. Thus
considering a small interval in the neighbourhood of the point charge, they
deduced,
\begin{equation}
F^\mu_{ret} = \frac{1}{2} \{ F^\mu_{ret} + F^\mu_{adv} \} + \frac{1}{2}
\{ F^\mu_{ret} - F^\mu_{adv} \}\label{e108}
\end{equation}
The left side of (\ref{e108}) is the usual causal field, while the right side has
two terms. The first of these is the time symmetric field while the second
can easily be identified with the Dirac field above and represents the sum
of the responses of the remaining charges calculated in the vicinity of the
said charge. Also here we encounter effects within the Compton scale (Cf. ref.\cite{r7}) of the rest of the universe.
We thus return to the concept from Quantum
Superstring theory, or more generally a theory based on relations like
(\ref{e1}) of extended particles and duality, a manifestation of holism.\\
5. One could argue that the non commutative relations (\ref{e1}) are an expression
of Quantum Mechanical spin. To put it briefly, for a spinning particle the non
commutativity arises when we go from canonical to covariant position variables.
Zakrzewsk\cite{r28} has shown that we have the Poisson bracket relation
$$\{x^j, x^k\} = \frac{1}{m^2}  R^{jk}, (c = 1),$$
where $R^{jk}$ is the spin. The passage to Quantum Theory then leads us back
to the relation (\ref{e1}).\\
Conversely it was shown that the relations (\ref{e1}) imply Quantum Mechanical
spin\cite{r25}. Another way of seeing this is to observe  that (\ref{e1}) implies
that $y = \alpha \hat p_y ,$ where $\alpha$ is a dimensional constant viz
$[T/M]$ and $\hat p_y$ is the analogue of the momentum, but with the Planck
constant replaced by $l^2$. So the spin is given by
$$| \vec r \times \vec p | \approx 2 x p_y \sim 2 \alpha^{-1} l^2 =
\frac{1}{2} \left(\frac{\hbar}{m^2c^2}\right)^{-1} \times \frac{h^2}{m^2c^2} =
\frac{\hbar}{2}$$
as required.

\end{document}